\def\kms.{\ifmmode $km~s$^{-1}\else km~s$^{-1}$\fi}

\documentstyle[11pt,aaspp4]{article}

\lefthead{Djorgovski et al.}
\righthead{Collapsed Cores in Globular Clusters}

\begin{document}

\title{The Eastern Arm of M83 Revisited:\\ High-Resolution Mapping 
of $^{12}$CO 1--0 Emission}

\author{Richard J. Rand}
\affil{Dept. of Physics and Astronomy, University of New Mexico, 800 Yale
Blvd, NE, Albuquerque, NM 87131}

\author{Steven D. Lord}
\affil{Infrared Processing and Analysis Center, MS 100-22, California Institute
of Technology, Pasadena, CA 91125}

\and

\author{James L. Higdon}
\affil{Kapteyn Astronomical Institute, University of Groningen, Postbus 800,
9700 AV Groningen, The Netherlands}



\begin{abstract}
We have used the Owens Valley Millimeter Array to map $^{12}$CO
(J=1-0) along a 3.5 kpc segment of M83's eastern spiral arm at
resolutions of $6.5\arcsec \times 3.5\arcsec$, $10\arcsec$, and
$16\arcsec$.  The CO emission in most of this segment lies along the
sharp dust lane demarking the inner edge of the spiral arm, but beyond
a certain point along the arm the emission shifts downstream from the
dust lane to become better aligned with the young stars seen in blue
and H$\beta$ images.  This morphology resembles that of the western
arm of M100.  Three possibilities, none of which is wholly
satisfactory, are considered to explain the deviation of the CO arm
from the dust lane: heating of the CO by UV radiation from young
stars, heating by low-energy cosmic rays, and a molecular medium
consisting of two (diffuse and dense) components which react
differently to the density wave.  Regardless, the question of what CO
emission traces along this spiral arm is a complicated one.  Masses
based on CO emission and the virial theorem for ten emission features
roughly agree and are in the range 1.5--16 $\times 10^6$ M$_{\sun}$.
These are lower than the masses of GMAs in M51, but the discrepancy is
probably due to the much higher linear resolution of these
observations.  Despite the uncertainty in what CO emission is tracing,
we do not require a conversion factor of CO brightness to H$_2$ column
density much different from the standard Galactic value if these
structures are bound.  Surprisingly, for the two fields where we can
compare with single-dish data, only 2--5\% of the single-dish flux is
seen in our observations.  A possible explanation is that M83 contains
much smoothly distributed molecular gas that is resolved out by the
interferometer.  Strong tangential streaming is observed where the arm
crosses the kinematic major axis of the galaxy, implying that the
shear becomes locally prograde in the arms.  The amplitude of the
tangential streaming is used along with a low-resolution single-dish
radial profile of CO emission to infer a very high gas surface density
of about 230 M$_{\sun}$ pc$^{-2}$ and an arm-interarm contrast greater
than 2.3 in the part of the arm near the major axis.  Using two
different criteria, we find that the gas at this location is well
above the threshold for gravitational instability -- much more clearly
so than in either M51 or M100.  This finding is consistent with the
unusually high H$\alpha$ surface brightness and star formation
efficiency in M83: star formation may be particularly active because
of strong gravitational instabilities.

\end{abstract}




%

\section{Introduction}

M83 is one of the brightest known galaxies in terms of molecular
emission [$I_{CO}$ = 64 K km s$^{-1}$ over the inner 2 kpc; Lord
(\markcite{} 1987)], and is the nearest ($D$ = 5 Mpc) barred spiral.
It is also notable for its large angular size ($D_{25}$ =
12.9$\arcmin$), its large and massive bar, and high star formation
rate ( 5 M$_{\sun}$ yr$^{-1}$; Kennicutt, Tamblyn, \& Congdon
\markcite{} 1994).  Other global properties are summarized in Table 1.

The eastern arm in particular has been popular for studies of star
formation and spiral density wave dynamics because it shows a clearly
delineated dust lane bordering a grand design spiral arm (Figure 3,
Plate 00, top left panel).  A series of detailed studies of this
region began with Allen et al. (\markcite{} 1986), who argued that
compression of the molecular gas in the spiral shock at the dust lane
is triggering star formation.  About 400 pc ($15\arcsec$) towards the
concave or outer edge of the arm (i.e. the downstream edge assuming a
trailing arm inside the corotation radius of the spiral density wave;
under this assumption, rotation proceeds clockwise), newly-formed
stars photodissociate the H$_2$ in GMCs and produce the HI ridge found
along the optical arm.

However, Wiklind (\markcite{} 1990), and Lord \& Kenney (\markcite{}
1991; hereafter LK) found evidence to the contrary: CO observations of
the southern part of this arm showed that the molecular emission does
not peak at the dust lane but instead downstream at the very sites
where Allen et al.\ would have a significant fraction of cloud H$_2$
mass converted to HI.  The results suggest that UV illumination could
highlight molecular clouds close to newly-formed stars by elevating
cloud temperatures, and thus that CO emission does not reflect the
true distribution of molecular gas.  However, LK argued that UV
heating of clouds in the arm alone is incapable of producing all of
the observed CO emission.  Rather, cloud compression in the optical
arm (rather than in the dust lane) is suggested.

A simpler picture can be drawn for M51, where the excellent
coincidence of the CO emission and the dust lanes throughout the
galaxy implies that CO is a good tracer of molecular gas column
density (Vogel, Kulkarni, \& Scoville \markcite{} 1988; Rand \&
Kulkarni \markcite{} 1990; Gruendl \markcite{} 1996).  The HI is
associated with the arms of HII regions, generally downstream from the
CO arms and dust lanes, and thus can be naturally explained as a
dissociation product (Tilanus \markcite{} 1990; Rand, Kulkarni, \&
Rice \markcite{} 1992).  In the western arm of another grand-design
spiral, M100, Rand (\markcite{} 1995) found that CO is coincident with
the dust lane at the beginning of the arm, but further along it
becomes better aligned with star forming regions (Rand \markcite{}
1995).
 
A possible explanation for such varying morphologies is that in
CO-luminous galaxies like these three, much of the diffuse ISM may be
in molecular form (Elmegreen \markcite{} 1993).  Collision fronts for
diffuse and dense gas in general need not occur at the same place in a
spiral arm, depending on the strength of the density wave (Elmegreen
\markcite{} 1988).  If the wave is sufficiently weak, only diffuse gas
is trapped at the well-organized shock front marked by the dust lane
on the inner edge of a spiral arm.  The dense clouds (i.e. GMCs),
orbiting quasi-ballistically in a manner similar to the stars, can
pass through the shock front and form a broad ridge centered on the
spiral arm.  The GMCs would be responsible for the CO emission.  In
this interpretation, the strong density wave in M51 is capable of
compressing both the diffuse and dense molecular components at the
shock front, accounting for the excellent coincidence of CO emission
and the dust lanes.  In M100, the density wave would have to weaken
sufficiently with distance along the spiral arm so that both molecular
components are trapped at the shock front at the beginning of the arm
but only the diffuse component further along the arm.  In this
explanation, the shock front at the location investigated by LK would
also be relatively weak.

Deutsch \& Allen (\markcite{} 1993) argue for a third possibility to
explain the CO emission downstream from the dust lane: low-energy {\it
cosmic ray heating} (see also Adler, Allen, \& Lo \markcite{} 1991;
Suchkov, Allen, \& Heckman \markcite{} 1993).  They have produced a
non-thermal radio continuum map of M83 (from a 20-cm map and an
extinction-corrected H$\beta$ image used to estimate the thermal
component) at $10\arcsec$ resolution which shows a ridge of emission
well aligned with the dust lane along most of the eastern arm but
shifts away from it in the southern part to become better aligned with
the small CO features found by LK.  They suggest that the mid-arm
ridge is bright in CO due to cosmic ray heating brought on by frequent
supernovae, and not due to a gas density enhancement.  They predict
that enhanced CO emission should generally occur along this ridge.
Testing their model against observations of M51, they argue that the
CO arms are coincident with the dust lanes primarily because the
non-thermal emission peaks there too (see also Tilanus et
al. \markcite{} 1988).  The fundamental correlation, then, is claimed
to be not between CO and dust but between CO and non-thermal emission.
If true, we are faced with the uncomfortable proposition that CO
emission largely traces the effects of cosmic-ray heating rather than
the molecular mass distribution, at least in some parts of spiral
arms.  This has obvious consequences for our understanding of such
topics as the growth of molecular clouds and star formation.

Determining which of these mechanisms -- UV heating, cosmic-ray
heating, and selective compression of dual molecular gas components --
is dominant in M83's arms has not been possible due to the lack of CO
maps of sufficient sensitivity, resolution, and coverage to provide
good observational tests.  Hence, one of the main goals of the more
extensive mapping of CO in the eastern arm presented here is to probe
this issue further.

A second reason to study this arm further is to understand the
dynamical process of star formation in spiral density wave
compressions.  For M51 (Rand \markcite{} 1993b) and M100 (Knapen et
al. \markcite{} 1996), it has been found that the efficiency of star
formation is higher in the arms than in the interarm regions.  One
possible reason is that the arms are conducive to large scale
gravitational instability in the gas, as discussed by (e.g.) Kennicutt
(\markcite{} 1989) and Elmegreen (\markcite{} 1994).  Applying
theoretical criteria from these studies to the observations indicates
(Rand \markcite{} 1993b; Rand \markcite{} 1995) that the gas surface
density in the spiral arms in both galaxies is indeed somewhat above
the threshold for the onset of instabilities, while interarm gas
hovers near the threshold values, at least in the parts of the
galaxies where arm and interarm surface densities can be determined by
the kinematic method described in \S 3.5.  These comparisons, of
course, depend on CO being a fair tracer of molecular gas,
encapsulated in a global relation between CO emission and H$_2$ column
density -- the $X$ factor.  If $X$ is lower by a factor of two or
three than the Galactic or \lq\lq standard\rq\rq\ value of $X_{Gal} =
2.8 \times 10^{20}\ {\rm mol\ cm^{-2}\ (K\ km\ s^{-1})^{-1}}$ in these
galaxies, as hinted at by some studies (Rand \markcite{} 1993a; Adler
et al. \markcite{} 1992; Nakai \& Kuno \markcite{} 1995; Rand
\markcite{} 1995), then the gas in the arms would be at about the threshold
surface density.  In M83 as well, there must be some doubt about the
$X$ factor given the uncertain interpretation of the CO emission as
discussed above.

Measures of star formation rates, surface densities, and global
efficiencies for the galaxies M83, M51, and M100 are shown in Table 2.
All molecular masses assume $X_{Gal}$ and are defined for the regions
listed in the footnote.  While these masses are lower limits because
the outer regions of the galaxies have not been mapped, it is unlikely
that a large fraction of their H$_2$ masses lies outside these areas.
The H$\alpha$ luminosity of M83 is comparable to that of M100 and M51,
but the surface density of H$\alpha$ emission is much higher in M83
and M51 than in M100.  Also, M83's global star formation efficiency,
defined observationally as $L_{H\alpha}$/$M_{H_2}$, is about 2.5 times
that of M51 and more than 3 times that of M100, subject to
uncertainties in $X$.  It is an important open question whether these
apparent variations in SFE can be related to levels of gravitational
instability in the molecular gas from one galaxy to the next.

To summarize, in this study we hope to further our understanding of CO
emission and the star formation process in galaxies by mapping a much
larger segment of the eastern arm of M83.  We would like to understand
in general the degree to which the CO emission reflects compression in
a density wave, cosmic ray heating, or UV heating from newly formed
stars.  Specifically, we focus on the mystery of the differences
between M83, M51 and M100.  Does the CO in M83 generally coincide
better with the dust lane, the young stars or the non-thermal
emission?  We also wish to understand how the density wave compression
raises the molecular gas surface density in M83, especially in
comparison with M51 and M100, and we will apply the aforementioned
criteria for gravitational instability to see whether the very high
surface density of star formation may be due to gas highly prone to
gravitational collapse.

\section{Observations}

Four fields along M83's eastern arm were imaged with the six-element
Owens Valley Radio Observatory (OVRO) Millimeter Array.  With
pointings separated by about $40\arcsec$ (primary-beam
$\theta_{FWHM}=65\arcsec$ at 115 GHz), there is significant overlap
between the fields (see the dashed circles in Figure 2).  The observed
region runs along the arm starting at the northern end of the central
bar, and extending about $2.5\arcmin$ (3.5 kpc) to the south. The
region was specifically chosen to cover the area studied by Deutsch \&
Allen (\markcite{} 1993) and the other works mentioned above.
Attention was paid to arrange the fields to sample the sharp dust lane
and to include the full width of the spiral arm.

Offsets of the pointing centers of the four fields from the center of
the galaxy are shown in Table 3.  The fields were observed between
December 1995 and June 1997 in three different configurations with
baselines ranging in length from 15 to 85 m, providing good uv
coverage for low-declination sources.  The quasar 1334--127 was used to
correct for temporal amplitude and phase variations.  The flux
standard for the maps was primarily determined through observations of
Neptune, while for some tracks the sources 3C273 and 3C345 were used.
These quasars were also used for bandpass calibration.  Spectral
coverage was provided by 32 1-MHz correlator channels resulting in a
resolution of 2.6 km s$^{-1}$ and an instantaneous bandwidth of 83.2 km
s$^{-1}$.  Naturally weighted channel maps separated in velocity by 5.2
km s$^{-1}$ were made and CLEANed using NRAO's\footnotemark{} AIPS
software package.

\footnotetext{The National Radio Astronomy Observatory is operated by
the Associated Universities, Inc. under cooperative agreement with the
National Science Foundation.}

The resulting synthesized beam FWHM was $6.5\arcsec \times 3.5\arcsec$
(P.A. = $-10^{\arcdeg}$), which at the assumed distance of 5 Mpc
corresponds to a linear resolution of 160 $\times$ 85 pc.  There were
slight variations in resolution from field to field.  Structures as
large as $\sim 30\arcsec$ (750 pc) should be well imaged.  The
1$\sigma$ map noise for each field in 5.2 km s$^{-1}$ channels is
listed in Table 3.  The four cubes were linearly mosaicked together
and corrected for primary beam attenuation using the task ``linmos''
in the MIRIAD package (Sault, Teuben, \& Wright \markcite{} 1995).
Primarily for the purpose of examining the velocity field, maps of
each field at about $10\arcsec$ resolution were made by tapering the
uv data (again there is a slight variation in resolution from field to
field), and a mosaic cube was formed from these.  A mosaic cube at
$16\arcsec$ resolution was also produced by convolving the
$10\arcsec$-resolution mosaic cube with a Gaussian to match the {\it
linear} resolution of the observations of M51 by Rand \& Kulkarni
(\markcite{} 1990).  Integrated line intensity and velocity field maps
were made by first masking emission free regions in the channel maps,
and then using only those pixels with signals exceeding a specified
($1.5-2\sigma$) noise level.


Although the map of non-thermal emission by Deutsch \& Allen
(\markcite{} 1993) provides a useful comparison with the CO data,
their original Very Large Array (VLA) 20-cm continuum map has
considerably lower resolution (10$\arcsec$) than our CO map.  In order
to compare the CO and 20-cm distributions at matched resolutions, we
made new {\em robust} weighted maps using the VLA B-array data
provided by J. Cowan (cf. Cowan, Roberts, $\&$ Branch \markcite{}
1994).  Setting the ``robust'' parameter to --0.2 in the AIPS task
IMAGR gave final CLEANed maps with synthesized beams nearly matching
the OVRO data (7.0$\arcsec \times 3.4\arcsec$), with only a
$\sim$10$\%$ increase in the map noise (40 $\mu$Jy beam$^{-1}$) over
natural weighting.  The wide range in baselines ensured that emission
structures as large as $\sim$100$\arcsec$ would still be well imaged.
Note that we have not attempted to remove a thermal component from
this map.

\section{Results}

\subsection{General Morphology and Comparison with Other Tracers}

Figure 1 shows the channel maps from the full-resolution cube for the
velocity range of 427.6--516.0 \kms.  The full-resolution map of total
CO intensity is shown in Figure 2.  Figure 3 shows overlays of CO
contours on other spiral tracers: (top left panel) a blue CCD image;
(top right) the 20-cm map; (bottom left) an H$\beta$ image from
Tilanus \& Allen (\markcite{} 1993; hereafter TA); and (bottom right) a
three-color image in which blue, red, and green represent CO, 20-cm,
and H$\beta$ emission, respectively.

\placefigure{fig1}

\placefigure{fig2}

Figure 2 clearly shows a segment of a molecular spiral arm, with much
substructure.  The relationship between the CO emission and the dust
lane revealed by the blue CCD image in Figure 3 is very complex.  At
the beginning of the mapped part of the arm, there is a noticeable
bifurcation of the dust lane at R.A. 13$^{\rm h}$ 34$^{\rm m}$
15$^{\rm s}$, Dec. $-29\arcdeg 36\arcmin 10\arcsec$.  These two lanes
appear to merge again at about R.A. 13$^{\rm h}$ 34$^{\rm m}$
18.5$^{\rm s}$, Dec. $-29\arcdeg 36\arcmin 30\arcsec$.  There are also
patches of extinction between these two lanes.  In this region, the CO
mainly coincides with the prominent outer dust lane, although faint
emission can be seen at the beginning of the inner dust lane as well.
South of where the lanes join again, there is good coincidence of the
CO emission and the dust lane.  However, at about Dec. $-29\arcdeg
37\arcmin$, the CO emission deviates from the main dust lane and
becomes more coincident with the young stellar clusters downstream,
although at least two CO clumps are still associated with the
prominent dust lane.  This part of the arm includes the region mapped
by LK and the current map confirms their finding of an offset while
revealing much more information about the CO-dust relationship.  We
will often refer to this part of the arm as the ``southern segment,''
and the part further to the north where CO and dust are more
coincident as the ``northern segment.''  To the degree that CO shows
general alignment with the dust lane at the beginning of the arm but
better alignment with star formation further along the arm, the
eastern arm of M83 resembles the western arm of M100 mapped by Rand
(\markcite{} 1995).  This general CO-dust morphology was predicted by LK.

Despite their prominence, the dust lanes in the southern segment do
not show substantial detectable CO emission.  However, by our
estimates this is not surprising.  From the excess extinction
(A$_{B}$ = 0.9) LK inferred a corresponding gas column density of
$\sim 5 \times 10^{21}$ mol cm$^{-2}$.  A cloud with this average
column density would have a CO brightness of $\sim$2 K km s$^{-1}$,
assuming X$_{Gal}$.  For a triangular line profile with a FWHM of 8 km
s$^{-1}$, the peak brightness would be 0.3 K.  Our 1$\sigma$
sensitivity in the channel maps is 0.2 K at full-resolution and 0.1 K
at $10\arcsec$ resolution.  Hence, a cloud of this column density,
reasonably well matched to the beam size, would be a marginal
detection in our maps.  Detectability is improved if the gas is
clumped.  We do see one feature (No. 10 in Figure 2) on the dust lane
in the southern segment, with an average column density of about
$2\times 10^{21}$ mol cm$^{-2}$.

As an aside, in Wiklind's (\markcite{} 1990) SEST map of part of the
eastern arm (overlapping our fields 2, 3, and 4) CO is again generally
associated with the optical arm.  However, this result by itself does
not rule out a faint narrow ridge of molecular gas along the dust
lane.  Such a ridge may be so smoothed out by the $43\arcsec$ beam
that it cannot be distinguished from the broader CO emission from the
optical arm.

So perhaps it is not so surprising that CO emission is difficult to
detect on this segment of the dust lane.  On the other hand, LK
estimate that the extinction in the dust lane in M51's northwest arm
is about half the above value, yet CO emission is readily detectable.
However, LK also conclude that M51's dust lane -- and by implication
the molecular distribution -- is clumpier than M83's, making it more
easily detectable interferometrically.  Indeed, the dust lane in the
northern segment of the M83 arm appears to be patchier than the
southern segment lane.  Additional uncertainty in the comparison is in
the relative vertical distributions of stars and dust in the arms of
the two galaxies -- there may be much hidden dust in some regions.

Figure 4 shows CO contours on the non-thermal radio emission map of
Deutsch \& Allen (\markcite{} 1993).  Comparison of the CO
distribution with the two radio continuum maps in Figures 3 and 4 give
somewhat different impressions, but both show a complicated spatial
relationship.  In Figure 4, a rather broad non-thermal arm can be
seen, surrounded by lower-level emission.  Much of the beginning of
the northern CO segment generally lies along the northern edge of the
non-thermal arm, but some CO emission extends northward into the
low-level non-thermal emission.  Further to the west and south, the
coincidence is much better.  Interestingly, at the top of the southern
segment where the CO arm shifts away from the dust lane towards the
young stellar clusters, the non-thermal arm shows the same behavior.
Further down this segment, the non-thermal arm tends to wander back
and forth across the CO arm.  One limitation of this non-thermal map
is that the thermal component was estimated from an H$\beta$ image.
If there is patchy extinction and the estimated thermal fraction is
large enough then the non-thermal map calculated in this way could
show small-scale differences from the true distribution of emission.
A non-thermal map produced from multi-frequency radio observations
should provide a more robust comparison.

\placefigure{fig4}

The 20-cm map in Figure 3 shows, in the northern segment, a few bright
peaks of emission along the top edge of the still-present broad arm.
These form a ridge somewhat coincident with the CO arm.  However,
there are several CO peaks which do not have associated continuum
peaks, and some CO emission still extends northward of this ridge.
Also, the three brightest 20-cm peaks coincide with HII regions (see
the three-color image), and may therefore have a substantial thermal
component.  Again, high-resolution multi-frequency radio observations
could resolve this issue.  Finally, there is little CO emission
associated with the lower-level broad arm of 20-cm emission, despite
the fact that this arm overlaps the southern bifurcation of the dust
lane.  In the southern segment two peaks of CO and 20-cm emission
coincide, but on scales close to the resolution the correlation is not
good.

Figure 3 shows that the relation between CO and H$\beta$ emission is
also rather complex, and again shows many similarities to M100.  In
the northern segment, CO is generally found at the inner edge of the
H$\beta$ arm.  In other words, there is no systematic offset of CO and
the most luminous HII regions as is the case in M51, but much more
H$\beta$ emission is found downstream of the CO than upstream, as seen
in the northern segment of the western arm of M100.  As explained by
Rand (\markcite{} 1995), this morphology is still consistent with
triggered star formation in a molecular compression (as long as CO
emission traces the compression), but if there is any time delay
between the compression of the gas and the onset of massive star
formation, it does not lead to a large CO-H$\beta$ offset.  This can
happen if the delay is rather short, if star formation is not
precisely sequenced along the arm, or if the velocity component
perpendicular to the arm of the gas flow vector is sufficiently small.
On smaller scales, patches of CO emission are sometimes coincident
with HII regions, sometimes not.  In the southern segment, the CO arm
is coincident with the middle of the H$\beta$ arm, which is delineated
by luminous HII regions, again very similar to the M100 case.  The
small-scale correlation of CO and H$\beta$ peaks is again poor.

A comparison of Figure 17 of TA with Figure 3 shows that there is no
clear systematic offset between M83's HI and CO arms.  Only in Field 2
(roughly between Declinations $-29\arcdeg 35\arcmin 45\arcsec$ and
$-29\arcdeg 36\arcmin 45\arcsec$) can it be said that HI is found
preferentially downstream of CO.  On the other hand, given the lack of
a clear CO-H$\beta$ offset even in the northern segment, it may still
be the case that the HI is produced by dissociation of H$_2$ by newly
formed stars.  The dissociation picture is difficult to establish when
spatial offsets are not clear, let alone the origin of the CO
emission.

\subsection{Fraction of Single-dish Flux Recovered}

Lord (\markcite{} 1987) measured a line flux of 1300 Jy km s$^{-1}$
with the FCRAO 14-m telescope near the center of our Field 1 (Table
1).  Within an area equal to the FCRAO beam (45 $\arcsec$) we detect
30 Jy km s$^{-1}$, or only about 2\% of the single-dish flux.  LK
similarly observed their field using the NRAO 12-m telescope and
measured a line flux of 357 Jy km s$^{-1}$.  Over the corresponding
single-dish beam area (65$\arcsec$) we measure 18 Jy km s$^{-1}$, or
again only a small fraction (5\%) of the total line flux.  These
recovered fractions are much lower than in interferometric
observations of other spirals: e.g. 35\% for OVRO observations of M51
(Rand \& Kulkarni \markcite{} 1990) and 55\% for BIMA observations of
M100 (Rand \markcite{} 1995).  A contributing factor may be the
combination of a small synthesized beam and the relative proximity of
M83, leading to a linear resolution of only $\sim$100 pc.  The 400-pc
and 600-pc linear resolutions of the M51 and M100 observations,
respectively, may be better matched to the width of the molecular
spiral arms, leading to a larger recovered fraction of the flux.

On the other hand, in two recent interferometric studies of another
nearby grand-design spiral M81 (Taylor \& Wilson \markcite{} 1998;
Brouillet et al. \markcite{} 1998), with beam sizes of about
$3\arcsec$ (45 pc for $D=3$ Mpc ) and $5\arcsec$ (75 pc) respectively,
a majority of the single-dish flux was recovered.  Perhaps in CO-rich
galaxies such as M83, M51, and M100, the molecular gas is relatively
widespread on scales larger than GMCs (such as GMA, spiral arm, and
galaxy-wide scales), while in a CO-poor galaxy such as M81, the
molecular gas is concentrated into relatively small, GMC-sized units
embedded in a sea of HI, making it possible to recover most of the
emission despite the spatial filtering of an interferometer.

\subsection{Properties of Discrete Emission Features}

The CO emission along the spiral arm is very clumpy, with many features
standing out as well-defined peaks, while other parts show more
smoothly distributed emission.  The features which are bright and
distinct enough to allow an accurate measurement of their fluxes,
sizes and linewidths are labeled in Figure 2.  We compute masses based on
CO flux and on the virial theorem for these features.  Of course, for the
former mass estimates, we assume that CO emission is a fair tracer of
molecular column density -- an assumption that may be questionable given
our lack of understanding of CO excitation in this galaxy.

Masses based on CO flux, $M_{CO}$, are calculated from the full-resolution
map using the following formula:
 
\begin{equation}
M_{CO} = 3.3 \times 10^5\ {F \over ({\rm Jy\ km\ s^{-1}})} ({ D \over 5\ {\rm Mpc}}
)^2
{X \over X_{Gal}},
\end{equation}
where $F$ is the integrated flux from the GMA, and $D$ is the
distance to M83.  Values of $M_{CO}$ are listed in Table 4 for $X=X_{Gal}$.

Uncertainties arise from the assumed distance, the conversion factor
adopted, and the flux calibration.  The distance is assumed to be 5
Mpc, but values as low as 3.7 (de Vaucouleurs \markcite{} 1979) and as
high as 8.9 Mpc (Sandage \& Tammann \markcite{} 1975) have been used.
Conservatively using these extremes as an indication of the
uncertainty in that quantity, the fractional uncertainty from flux
calibration and distance considerations alone is 1.0.  We will return
to the question of the value of $X$ after comparing these masses with
virial masses.

We calculate the virial mass in the same way as Rand (\markcite{} 1995):
\begin{equation}
M_{vir} = 550d(\sigma_{1d})^2 M_{\sun} 
\end{equation}
where $d$ is the deconvolved diameter and $\sigma_{1d}$ is the 1-d
velocity dispersion.  The latter is estimated from the measured FWHM
of the line profile averaged over the area of the feature.  Most of
the features are resolved and we use the geometric mean of their sizes
to calculate $d$.  The formal uncertainties in $M_{vir}$ are due to
the distance uncertainty and the approximately 2 km s$^{-1}$
uncertainty in the estimated FWHM of the spectra.  In general, the
virial masses are quite comparable to the flux-based masses given the
large uncertainties, with the main exceptions being features 8, 9, and
10.

The range of masses extends from the equivalent of the largest
Galactic GMCs (e.g. Scoville \& Sanders \markcite{} 1987) to the
smaller GMAs of M51 (Rand \& Kulkarni \markcite{} 1990).  If these
features are really bound units, then the assumed value of $X$ must be
reasonable.  However, from recent studies of other nearby spirals, one
could make an argument that the appropriate value of $X$ may be two or
three times lower than assumed.  M83 has a high oxygen abundance: six
HII regions in M83 have been studied by Dufour et al. (\markcite{}
1980), with three of these in a range of galactocentric radii similar
to that of the CO emission.  They have abundances of 12 + log(O/H) =
9.64, 9.54, and 9.43.  For M51, a galaxy with an oxygen abundance as
high as that of M83 (Dufour et al. \markcite{} 1980), a value of $X$
three times lower than the standard Galactic value is hinted at by
Rand (\markcite{} 1993a), Adler et al. (\markcite{} 1992), and Nakai
\& Kuno (\markcite{} 1995).  Wilson (\markcite{} 1995) finds from a
comparison of CO-based and virial masses of GMCs in five Local Group
galaxies (IC 10, M31, M33, NGC 6822, and the SMC) that $X$ increases
by a factor of 4.6 as the oxygen abundance decreases by a factor of
10.  While Wilson (\markcite{} 1995) does not include galaxies with
abundances as high as M83 or M51, extrapolation of her relationship
would also indicate a value of $X$ about three times below the
Galactic value.  Obviously, such an extrapolation is dangerous, and
theoretically at least, while a dependence of $X$ on abundance is
expected, it is predicted to be weak for abundances above solar
(Maloney \& Black \markcite{} 1989).  Adoption of the lower value of
$X$ would cause the majority of the features found here to be unbound.
While possible, this conclusion would challenge the idea, developed
from observations of several nearby spirals, that galaxies with high
molecular contents and strong density wave compressions should be able
to grow larger bound units.  At this point, then, there is currently
no compelling reason to assume a significantly lower value of $X$.

Apart from their boundedness, why do we not see features as massive as
in M51?  It is no doubt the case that the typical size of features
seen is an effect of resolution.  Because of M83's relative proximity
and the high angular resolution of the observations, we tend to see
smaller structures than in M51 [note that even larger features were
found in M100 by Rand (\markcite{} 1995), where the linear resolution
was about 600 pc].  Discrete features at this linear resolution (160
$\times$ 85 pc) will blend together when observed at the 370-pc
resolution of the M51 observations of Rand \& Kulkarni (\markcite{}
1990).  To demonstrate this, we examine a total-intensity map made
from the $16\arcsec$-resolution cube, which was created to match the
linear resolution of the M51 observations.  It shows only four
discrete clumps, each an association of two or three of the features
observed at full resolution.  Their flux-based masses, using
$X_{Gal}$, are now in the range $1.1-4.3
\times 10^7$ M$_{\sun}$ -- more comparable to the typical $3\times
10^7$ M$_{\sun}$ mass of the GMAs in M51.

Interestingly, though, the virial masses are generally much higher
than the flux-based masses -- in the range $5.9-27 \times 10^7$
M$_{\sun}$ -- mainly due to the rather large line-widths (two are
strictly upper limits due to being spatially unresolved, but from the
full-resolution map it can be inferred that their sizes should be
comparable to the $16\arcsec$ beam).  Hence, at this resolution, they
appear as unbound GMAs.  A lower $X$ would worsen the discrepancy
[whether the arm GMAs in M51 are unbound is complicated by the fact
that many are unresolved (Rand \& Kulkarni \markcite{} 1990) as well
as the uncertain value of $X$].

\subsection{Kinematics and Streaming Motions}

It was concluded by TA from analysis of VLA HI and Fabry-Perot
H$\beta$ velocity fields that density wave induced streaming motions
are much weaker in M83 than in M51.  After subtraction of a rotation
curve (derived from the HI data) which rises linearly to 185 km
s$^{-1}$ at $R=2.5\arcmin$ and is flat beyond this radius, they found at
most a suggestion of a 15 km s$^{-1}$ (in the plane of the sky)
velocity gradient across the arms in the H$\beta$ velocity field.

The general features of the CO velocity field are most clearly
revealed by the $10\arcsec$ velocity field, shown in Figure 5.  The
lines indicate the major and minor axes derived by TA.  The detected
CO emission is entirely on the approaching (northeast) side of the
galaxy.  The southeast half of the galaxy is the far side under
the assumptions in \S I.  Our fields include a location where the arms
cross the major axis, allowing tangential streaming to be examined.
The southern end of the detected arm is close to the minor axis, but
the emission is very weak and covers a small across-arm width.  Hence,
an examination of radial streaming is much more difficult.

\placefigure{fig5}

We first note that there is in general a sharp decrease in observed
velocity (translating to an increase in tangential velocity) across
the arm near the major axis.  The rate of decrease is even higher in
the full-resolution velocity field.  This gradient could be due to
either tangential streaming or a rising rotation curve.  TA did in
fact derive a rising curve from their HI data, but the slope is much
too small to explain the gradient.  On the approaching side, they find
$V_{rot}$ increases in this radial range by about 10 km s$^{-1}$ over
$20\arcsec$ (the approximate width of the detected CO arm near the
major axis), or 4 km s$^{-1}$ in the plane of the sky.  The observed
gradient in Figure 5 is about 20--40 km s$^{-1}$ .  Hence, the CO
shows rather strong tangential streaming in the sense expected (e.g.
Roberts \& Stewart \markcite{} 1987) for gas flowing through a density wave
compression.  The gradient in the CO data varies somewhat along the
arm, but is typically 120 km s$^{-1}$ kpc$^{-1}$ and 85 km s$^{-1}$
kpc$^{-1}$ in the full- and $10\arcsec$-resolution CO velocity fields,
respectively.  The different gradients indicate that resolution is an
issue, and it is possible that an even larger value would be observed
at higher resolution.  In contrast, TA found no evidence for strong
streaming in their $12\arcsec$-resolution H$\beta$ velocity field.
Their data seem to be more dominated by velocity irregularities, which
may mask more regular streaming motions.

At locations close to the minor axis on the far side, the radial
inflow of gas as it passes through the compression should be
manifested as an observed velocity decrease from the front to the back
of the arm.  No such trend is clear in Figure 5, but for reasons
stated above, one may not expect such a gradient to be seen in our
data.

\subsection{Surface Density Contrasts and Gravitational Instability}

The tangential velocity gradient can be used to show how the usual
shear associated with a flat rotation curve is modified in a spiral
arm (e.g. Elmegreen \markcite{} 1994; Rand \markcite{} 1993b,
\markcite{} 1995).  The rate of shear is given by the Oort $A$
constant:
\begin{equation}
A = 0.5({{v}\over R} - {{\partial v}\over {\partial R}})
\end{equation}
The rotation speed at $R=2.3$ kpc is 140 km s$^{-1}$ (TA).  For the
case of no streaming and using the slightly rising rotation curve of
TA (see \S 3.3), $A=20$ km s$^{-1}$ kpc$^{-1}$.  $A$ = 0 km s$^{-1}$
kpc$^{-1}$ corresponds to local solid-body rotation.  For the
tangential velocity gradient of 120 km s$^{-1}$ kpc$^{-1}$ measured
from the full-resolution velocity field, $A$ becomes --30 km s$^{-1}$
kpc$^{-1}$.  The negative value indicates that the shear in the arms
is prograde: gas at the back of the arms has a higher orbital angular
frequency than gas at the front.  This situation should reverse itself
in the interarms (as observed in M51; Rand \markcite{} 1993b), but we
cannot test this because we have not detected any interarm CO
emission.

The tangential streaming, if it is measurable, can also be used to
estimate the spiral arm surface density enhancement in the molecular
gas.  This dynamical method, based on Balbus \& Cowie (\markcite{}
1985), has been applied to the grand-design spirals M51 (Rand
\markcite{} 1993b) and M100 (Rand \markcite{} 1995), and avoids the
issue of whether CO emission traces molecular mass in the same way in
both arm and interarm regions.  The method assumes a tightly wound
spiral in which mass and angular momentum are conserved as the gas
flows in and out of the arms.  In this case, the local \lq\lq
effective\rq\rq\ epicyclic frequency (different in the arm, interarm,
and axisymmetric case because of its dependence on streaming
motions) of the gas varies with the local surface density, and it can
be shown that the ratio of arm to axisymmetrically averaged surface
densities is given by:
\begin{equation}
{\Sigma_{arm}\over \Sigma_{axi}} = 
{\kappa^2_{arm}/\Omega_{arm}\over \kappa^2_{axi}/\Omega_{axi}} = 
{{1 + {R\over v}({\partial v\over \partial R})_{arm}}\over
{1 + {R\over v}({\partial v\over \partial R})_{axi}}}
\end{equation}
where $\kappa$ is the effective epicyclic frequency, and $v$ is the
tangential speed at galactocentric radius $R$.  Note that this method
is independent of errors in the distance to the galaxy or its
inclination.  Using the above gradient of 120 km s$^{-1}$ kpc$^{-1}$
and the rotation curve of TA at this radius, ${\Sigma_{arm}\over
\Sigma_{axi}} \approx 2.3$. Lord (\markcite{} 1987) found that the
azimuthally averaged molecular surface density at $R=2.3$ kpc,
assuming $X=X_{Gal}$, is about 100 M$_{\sun}$ pc$^{-2}$.  The
molecular surface density in this part of the northern segment of the
arm is then about 230 M$_{\sun}$ pc$^{-2}$.

The arm-interarm contrast will be greater than 2.3, but we cannot
estimate it without a determination of interarm streaming gradients.
However, it is interesting that Wiklind (\markcite{} 1990), using SEST
CO 1--0 and 2--1 data and a model with a composite of optically thin
and thick gas, derived a contrast somewhat further down the arm of at
least 4.5 on $45\arcsec$ scales.  These two independent methods
suggest that strong compression of the molecular gas occurs in M83 at
this radius.  In comparison, an arm-interarm contrast of 4 and an
arm-axisymmetric contrast of 2 were found for M51 at around $R=3$ kpc,
while values of 2--3 and 1.5 were found for M100 near $R=5$ kpc.  The
derived arm surface density is somewhat higher than in M51 (140 and
170 M$_{\sun}$ pc$^{-2}$ for the two arms) and much higher than in
M100 (60 M$_{\sun}$ pc$^{-2}$ in the northwest arm).

Finally, following Rand (\markcite{} 1995), we use two different
criteria for assessing whether the gas in the arms near the major axis
crossing should be gravitationally unstable.  The first is based on
observations: Kennicutt (\markcite{} 1989) finds that significant star
formation in spirals occurs only above a surface density threshold
related to the Toomre criterion for gravitational instabilities in a
uniform gas disk (Safronov \markcite{} 1960; Toomre \markcite{} 1964;
Quirk \markcite{} 1972):
\begin{equation}
\Sigma_c = {{\kappa \gamma^{1/2} c} \over {\pi G}}
\end{equation}
where $\Sigma_c$ is the threshold surface density, $\gamma$ is the
ratio of specific heats, here chosen to be 0.5 ({\it e.g.}  Elmegreen
\markcite{} 1992), $c$ is the one-dimensional gas velocity dispersion, and
$\kappa$ is the effective epicyclic frequency.

We use the value of $\kappa$ used to evaluate Eq. (4) for the arm and
$c=8$ km s$^{-1}$, the mean for the emission features.  We find
$\Sigma_c = 57$ M$_{\sun}$ pc$^{-2}$, well below the arm surface
density derived above.  The most uncertain observational input into
this comparison is probably the value of $X$, which would have to be
four times lower than assumed to make the gas in the arms marginally
stable.  The agreement of flux-based and virial masses for the
structures along the arms suggests that this is unlikely if the
structures have properties similar to Galactic GMCs.

The second criterion comes from a theoretical treatment of
gravitational collapse in a shearing, magnetized spiral arm by
Elmegreen (\markcite{} 1994), and was applied to M100 by Rand
(\markcite{} 1995).  The criterion describes whether an azimuthal
(along the arm) instability can collapse before it emerges from the
arm.  We use the same form as did Rand (\markcite{} 1995):

\begin{equation}
C = 0.5\,
Q_0^{-1.5}(\rho_s/\rho_0)^{1.75}(1-e^{-\chi})^{1.5}\Omega/(\Omega -
\Omega _P) > 1
\end{equation}
If $C>1$, there is sufficient time for collapse.  $Q_0$ is essentially
the Toomre criterion for the axisymmetric case, and is 0.5 for the
sound speed, rotation curve and axisymmetric surface density at
$R=2.3$ kpc used above. $\rho_s$ is the gas density in the arms and
$\rho_0$ is the axisymmetrically averaged gas density.  We assume that
the gas scale height does not change in the arms so that
$\rho_s/\rho_0 = \Sigma_s/\Sigma_0$.  The parameter $\chi =
(2G\mu_s)^{1/2}/c$, where $\mu_s$ is the gas mass per unit length
along the arm.  The latter quantity is the gas surface density
multiplied by the arm width.  Since the OVRO observations rapidly lose
sensitivity to emission on scales of $30\arcsec$ or larger, it is
likely that the arm is broader than observed.  An upper limit comes
from the fact that Wiklind (\markcite{} 1990) did not resolve the arm
with a $43\arcsec$ beam.  We set the arm width to $30\arcsec$, but in
reality any reasonable choice of arm width results in $e^{-\chi}<<1$.
For the pattern speed, we adopt the value of Kenney \& Lord
(\markcite{} 1991) of 51 km s$^{-1}$ kpc$^{-1}$, derived by assuming
that corotation of the bar is at 1.4 times the bar radius, or
$145\arcsec$.  While it is not clear that the bar and spiral should
have the same pattern speed (Sellwood \& Sparke \markcite{} 1988), the
spiral arms do begin at the ends of the bar, as would be expected if
arms and bar rotated as one pattern.  At $R=145\arcsec$, some of the
characteristics of corotation can be seen (Kenney \& Lord \markcite{}
1991): the HII regions in the eastern arm cross from the outside to
the inside edge of the arm, while in the western arm, the dust lane
crosses the arm and the density of HII regions decreases
significantly.

Using these values we derive $C=34$, which certainly implies
gravitational instability.  Can making reasonable changes to the
parameters bring it below unity?  The number is very insensitive to
the value of $\chi$.  Also, the velocity dispersion is reasonably well
determined and is unlikely to contribute much uncertainty to the
calculation.  Uncertainty in the arm surface density has a much more
dramatic effect on $C$ through $Q_0$: e.g. doubling the conversion
factor, $X$, lowers $C$ by a factor of 2.8.  Halving the pattern speed
(although there is no observational reason to do so) brings it down by
a factor of 3.7.  Changing the inclination to 14$\arcdeg$ will
increase $\kappa$, and thus $Q_0$, by a factor of 1.7, lowering $C$ by
a factor of 2.  Finally, depending on the balance of the velocity
dispersion and the gravity in the arms, the assumption of an
unchanging scale height as the gas passes through the arms may not be
true.  If the scale height were doubled in the arms (presumably due to
stirring by star formation activity), $C$ would decrease by 3.4. Taken
together, these four changes would lower $C$ to 0.5, but this
combination of circumstances is very unlikely and we conclude that the
gas is indeed gravitationally unstable in the arms at around this
galactocentric radius.

Although the comparison cannot be made for a broad range of
galactocentric radii, it is interesting that these instability
criteria are much more easily satisfied in M83 than in M51 or M100,
suggesting that gas is more prone to gravitational collapse in this
galaxy.  As discussed in \S 1, M83 also has the highest surface
density of star formation and a global star formation efficiency
about three times as high as in M51, subject to uncertainties in $X$.
The results are therefore suggestive of a link between star formation
activity and efficiency and the degree to which the gas is
gravitationally unstable to collapse.

\section{Discussion -- The CO Spiral Arm Morphology}

The CO morphology in relation to the other spiral tracers raises as
many questions as it answers.  Neither the dust lane nor the young
stars (as seen in a blue image) nor the non-thermal emission shows a
very good coincidence with the CO emission over the entire mapped
length of the arm.  The similarity in spatial relationships with the
western arm of M100 mentioned above suggests that there may be a
common physical process responsible for this state of affairs.

The three possible explanations we consider for this morphology -- UV
heating, cosmic ray heating, and a two-component molecular phase --
have their pros and cons, and show the need for further observations
and theoretical work:

1) UV heating.  One can argue that the CO-dust correlation is good in
the northern segment because the dust lane is sufficiently clumpy,
allowing the emission to be detected interferometrically.  The
southern dust lane is too smooth, and so CO is seen mostly near star
forming regions due to a combination of sufficient column density and
heating by stars.  In both segments, most of the brightest HII regions
are associated with the CO arm.  The small-scale CO-H$\beta$
correlation need not be perfect because {\bf a)} some CO emission
features may trace gas which has not yet formed many stars, {\bf b)}
some H$\beta$ peaks may represent regions where massive stars have
dispersed or dissociated the molecular gas, and {\bf c)} H$\beta$
should not trace star formation perfectly because of patchy
extinction.  The brighter HII regions in M83 indicate that stellar
heating is more important than in M51.  This scenario may not work so
well in M100 since the bright end of the HII region luminosity
function (Knapen \markcite{} 1997) much more closely resembles that of
M51 (Rand \markcite{} 1992) than that of M83 (Rumstay \& Kaufman
\markcite{} 1983; Kennicutt, Edgar, \& Hodge \markcite{} 1989).  It
should be remembered, however, that LK concluded that stellar heating
was insufficient to explain the CO morphology; that the southern ridge
must represent the peak of the molecular gas distribution.  Also, the
agreement of virial and flux-based masses for most of the emission
features, if they are bound, suggests that CO emission traces
molecular gas column density reasonably well and thus no strong
additional source of heating is indicated.

2) Low-energy cosmic ray heating.  For this mechanism to apply, the
emission should correlate best with the non-thermal radio continuum.
In favor of this explanation is the shift of both the CO and 20-cm
continuum arms away from the dust lane in the southern segment.  On
small scales however, the correlation is rather poor, and possible
thermal contamination in the high-resolution 20-cm map hampers the
comparison.  Again, the mass estimates for the emission features do
not suggest a strong source of additional heating.  A high-resolution
thermal-nonthermal separation using maps at 3.6, 6, and 20 cm rather
than an H$\beta$ image and a 20-cm map may be more robust and may
change the small-scale correlation.  The same should be done for M100.
Also, the distribution of low-energy cosmic rays may not be fully
reflected in the distribution of synchrotron emission.

3) Two-component molecular phase.  In this explanation, we are seeing
the reaction of such a medium to the density wave, as discussed in \S
1.  This requires that the density wave response weakens with distance
along the arm appropriately so that both diffuse and dense components
are trapped at the shock front in the northern segment, but only the
diffuse component in the southern segment.  The same scenario would
apply for M100, while in M51, the compression is strong enough over
the large part of the disk mapped in CO to trap both components along
the dust lanes.  The very low fraction of single-dish flux detected in
our fields is at least consistent with a prominent diffuse molecular
medium in M83.  Against this scenario, perhaps, is that the streaming
motions and inferred surface density of molecular gas in the arms are
similarly high in M83 (at least in the northern segment) and M51, and
lower in M100.  Does the compression weaken sufficiently in the
southern arm segment of M83 to allow the dense component to pass
through the shock?  Unfortunately, we cannot measure streaming motions
there with these data.  Further progress on this issue would be made
from a comparison of arm-interarm contrasts in near-IR observations --
assuming that the contribution of younger stars can be understood and
removed -- in M83, M51, and M100 as a function of galactocentric
radius, and more sensitive kinematic measurements.

There is evidence for the existence of widespread diffuse molecular
gas in the Milky Way (Polk et al. \markcite{} 1988) and nearby spirals
(Young \& Sanders \markcite{} 1986), and the observations of M33 by
Wilson \& Walker (\markcite{} 1994) suggest how this diffuse gas can
be largely missed in interferometric observations.  Clearly, the
detectability of diffuse gas with interferometers should depend on its
column density, CO emissivity and the linear beam-size.  Its behavior
in a spiral density wave compression relative to denser clouds could
be explored further in simulations of the reaction of a two-component
medium to density waves of various strengths.  The dense-diffuse
balance will depend on the ease of dense cloud creation, shredding of
dense clouds by the radiative and mechanical energy of star formation
(e.g. Elmegreen \markcite{} 1992) and perhaps in the density wave
compression, itself a possible source of heating (Thomasson, Donner,
\& Elmegreen \markcite{} 1991).  One would like to understand the
life-cycle of dense and diffuse gas as it passes in and out of a
spiral arm.

In all three scenarios, more could be gained by mapping the other arms
of M83 and M100 in CO emission, and comparing spiral tracers.  Of
course, some combination of these mechanisms may also be at work.
Also, any theory of CO heating must explain why flux-based and virial
mass estimates agree for most of the emission features discussed in \S
3.3, assuming they are bound.

Finally, a more realistic estimate of the distribution of extinction
would help in interpreting the CO emission.  The visual appearance of
dust lanes is due not only to the dust column density but also the
relative distribution of dust and stars.  In the star-forming part of
the arm, for example, it is possible that enough new stars sit above
the dust layer to give the appearance that there is much less dust
than is actually present.  Trewhella (\markcite{} 1997) shows that by using
multi-band optical and near-infrared photometry along with IRAS or ISO
maps of infrared emission, the inferred distribution of starlight and
extinction can be very different from the impression given by, say, a
blue image.  Applied to NGC 6946, his technique reveals a smooth,
two-armed extinction-corrected spiral pattern in blue light, in
contrast to the patchy, four-armed pattern familiar from uncorrected
blue images.  A similar analysis for M83 could be very enlightening.

\section{Conclusions}

The main conclusions from this study are:

1) We have detected molecular spiral structure along the eastern arm
of M83.  The northern part of the molecular arm shows reasonably good
coincidence with the dust lane, while the southern part is offset
downstream from the dustlane and shows better coincidence with the arm
of young stars, confirming the results of LK.  A map of non-thermal
emission shows a similar behavior, although the small- scale
correspondence with CO emission is sometimes poor.  There is little
evidence of a clear CO-H$\beta$ offset, but the fact that the CO
emission in the northern arm segment lies at the front of the H$\beta$
arm is still evidence for triggering of star formation by the density
wave compression.  The CO-dust-star formation morphology is similar to
that in the western arm of M100, but is different from that of M51, where
the CO-dust coincidence is excellent.

2) Three scenarios have been examined to to explain the relationship
of CO emission with other spiral tracers.  First, it may be that CO is
detectable in the northern segment because the dust lane is
sufficiently clumpy there, while a ridge of bright HII regions may
contribute to CO excitation.  In the southern segment, the dust lane
is rather smooth and CO is seen in the star forming arm due to heating
provided by young stars.  Second, CO emission may trace the
distribution of low-energy cosmic rays if they are responsible for
heating CO molecules.  Third, the morphology may be due to the
reaction of a two-component molecular medium to the spiral density
wave.  None of these explanations is completely satisfactory given
these new data.  Further observations and theoretical work should help
to distinguish between the possibilities.

3) An unusually low fraction of 2--5\% of the single-dish flux is recovered in
the interferometric maps.  This may be due to the proximity of the galaxy
combined with a prominent, smoothly distributed diffuse molecular medium which
is completely missed in these observations.

4) Emission features have masses ranging from the largest GMCs in the Milky Way
to the GMAs of M51.  Masses based on CO flux and virial masses roughly agree,
suggesting that, if these features are bound, the standard Galactic value of
$X$ is reasonable.  If true, this result must be accounted for by any theory
of the heating of the CO.

5) Strong tangential streaming, consistent with passage through a
density wave compression, is observed where the molecular arm crosses
the major axis.  From the amplitude of the streaming, the enhancement
in surface density of the arm over the disk average at that radius is
about 2.3, and the gas surface density in the arms is about 230
M$_{\sun}$ pc$^{-2}$.  The arm-interarm contrast will be greater than
2.3.  Two criteria for large-scale gravitational instabilities in the
arms are satisfied much more easily in M83 than in M51 or M100.
Enhanced gravitational collapse may be responsible for the relatively
high surface density and global efficiency of star formation in M83.

\acknowledgments

We are very grateful to J. Cowan for providing the 20-cm uv data,
E. Deutsch for the $10\arcsec$-resolution non-thermal map, and
R. Tilanus for the H$\beta$ image,

\begin{figure}
\caption{Channel maps from the $6.5 \times 3.5\arcsec$ resolution cube at
5.2 km s$^{-1}$ spacing.  Contour levels are --0.18, 0.18, 0.36, 0.54, and 0.72
Jy beam$^{-1}$.  The center velocity is indicated in the upper left
corner of each panel.\label{fig1}}
\end{figure}

\begin{figure}
\caption{Contours of total CO emission in M83 at $6.5\arcsec \times
3.5\arcsec$ resolution.  Contour levels are 2.6, 5.2, 7.8, 10.4, and 13
Jy (beam)$^{-1}$ km s$^{-1}$.  Numbers indicate emission features
discussed in \S 3.3.  The dashed outline indicates the half-power
points of the mosaicked primary beam response.  The FWHM beam size is
shown in the lower left corner.\label{fig2}}
\end{figure}

\begin{figure}
\caption{Top left: contours of total CO emission in M83 at $6.5\arcsec \times
3.5\arcsec$ resolution overlaid on a color representation of a blue
CCD image.  Top right: contours of total CO emission overlaid on a
color representation of the 20-cm continuum map at resolution matched
to the CO map.  Bottom left: contours of total CO emission overlaid on
the H$\beta$ image of Tilanus \& Allen (1993) at $4\arcsec$
resolution.  The truncation at the northeast edge of the H$\beta$
image marks the edge of the field of view of the observation.  Bottom
right: a three-color representation of the spiral tracers CO (blue),
20-cm continuum (red), and H$\beta$ (green).  CO contour levels are as
in Figure 2.  \label{fig3}}
\end{figure}

\begin{figure}
\caption{Contours of total CO emission at $6.5\arcsec \times
3.5\arcsec$ resolution overlaid on a grey-scale representation of
the map of non-thermal emission of Deutsch \& Allen (1993).  CO
contour levels as in Figure 2.\label{fig4}}
\end{figure}

\begin{figure}
\caption{Velocity field (first-moment map) from the $10\arcsec$-resolution
cube of CO emission shown as grey-scale with contours.  Darker shading
corresponds to higher velocities.  Velocity contours are from 430 km
s$^{-1}$ to 500 km s$^{-1}$ in steps of 5 km s$^{-1}$.  The 450, 470,
and 490 km s$^{-1}$ contours are shown in white.  The solid and dashed
lines represent the major and minor axes of the galaxy, respectively.
The beam-size is shown in the lower left corner.\label{fig5}}
\end{figure}
\newpage

\begin{table}[htb]
\begin{center}
\caption{General Parameters of M83}
\begin{tabular}{lll}
\tableline
 & & Reference \\
\tableline
 Type & SAB(s)bc & de Vaucouleurs et al. (1976) \\
 Right Ascension\tablenotemark{a}\ (1950.0) & $13^{\rm h}\, 34^{\rm m}\, 11.55^{\rm s}$ & \\
 Declination\tablenotemark{a}\ (1950.0) & $-29\arcdeg\, 36\arcmin\, 42\arcsec.2$ & Cowan et al. (1994) \\
 Heliocentric Systemic Velocity & 505 km s$^{-1}$ & Comte (1981) \\
 Distance & 5.0 Mpc & Kennicutt (1988) \\
 Linear Scale & $1\arcsec$ = 24 pc & \\
 Inclination & 24$\arcdeg$ & Talbot et al. (1979) \\
 Position Angle of Major Axis & 225$\arcdeg$ & Talbot et al. (1979) \\
 HI Mass\tablenotemark{b} & $1.0 \times 10^{9}$ M$_{\sun}$ & Tilanus \& Allen (1993) \\
 H$_2$ Mass within $R=115\arcsec$\tablenotemark{b,d}&  $5 \times 10^{9}$ M$_{\sun}$ &  Lord (1987)\\
 H$\alpha$ Luminosity\tablenotemark{b,c} & $3.7 \times 10^{41}$ erg s$^{-1}$ & Kennicutt, Tamblyn, \& Congdon (1994) \\
 FIR Luminosity\tablenotemark{b} & $1.3 \times 10^{10} L\sun$ & Rice et al. (1988)\\
\tableline
\end{tabular}
\end{center}
\tablenotetext{a}{Central radio continuum source}
\tablenotetext{b}{Scaled to $D=5$ Mpc}
\tablenotetext{c}{An extinction correction has been made by Kennicutt et al. (1994)}
\tablenotetext{d}{Assumes $X=2.8 \times 10^{20}\ {\rm mol\ cm^{-2}\ (K\ km\ s^{-1})^{-1}}$}
\end{table}
\newpage

\begin{table}[htb]
\begin{center}
\caption{Star Forming Properties of M83, M51, and M100}
\begin{tabular}{lllll}
\tableline
 & M83 & M51 & M100 & References \\
\tableline
Distance (Mpc) & 5.0 & 9.6 & 17.1 \\
$L_{H\alpha}$(10$^{41}$ erg s$^{-1}$)\tablenotemark{a} & 4.9 & 3.6 & 3.6 &  (1) \\
$M_{HI}$ (10$^{9}$ M$_{\sun}$) & 1 & 5 & 4 & (2) (3) (4) \\
$M_{H_2}$ (10$^{9}$ M$_{\sun}$)\tablenotemark{b} & 5 & 16 & 16 & (5) (6) (7) \\
$L_{H\alpha}$/$D_{25}^2$ (10$^{38}$ erg s$^{-1}$ kpc$^{-2}$)\tablenotemark{c} & 3.6 & 2.6 & 1.0 & \\
$L_{H\alpha}$/$M_{H_2}$ (10$^{31}$ erg s$^{-1}$ M$_{\sun}^{-1}$) & 7.2 & 3.1 & 2.5 & \\
$L_{H\alpha}$/$M_{HI + H_2}$ (10$^{31}$ erg s$^{-1}$ M$_{\sun}^{-1}$) & 6.0 & 2.4 & 3.1 & \\
\tableline
\end{tabular}
\end{center}
\tablenotetext{a}{An extinction correction has been made (see Kennicutt et al. 1994)}
\tablenotetext{b}{The same value of $X$ has been used for all galaxies (see text).  The H$_2$ masses
are for the inner $115\arcsec$, $170\arcsec$, and $200\arcsec$ for M83, M51, and M100, respectively}
\tablenotetext{c}{Isophotal diameters from de Vaucouleurs et al. (1991), corrected for inclination}
\tablerefs{(1) Kennicutt et al. 1994; (2) Tilanus \& Allen 1983; (3) Rots 1980;
(4) Knapen et al. 1993 (5) Lord 1987; (6) Kuno et al. (1995); (7) Sempere, \& Garcia-Burillo (1997).}
\end{table}
\newpage

\begin{table}[htb]
\begin{center}
\caption{Offsets from Nucleus and Map Noise Levels of Observed Fields}
\begin{tabular}{llll}
\tableline
Field & R.A. offset & Dec. offset & 1$\sigma$ rms\tablenotemark{a} \\
      & (arcmin) & (arcmin) & (mJy beam$^{-1}$) \\
\tableline
1 & --1.07 & \ \ 0.79 & 68 \\
2 & --1.54 & \ \ 0.34 & 57 \\
3 & --1.59 & --0.33 & 52 \\
4 & --1.50 & --0.91 & 45 \\
\tableline
\end{tabular}
\end{center}
\tablenotetext{a}{For 5.2 km s$^{-1}$ channels.  1 Jy (beam)$^{-1}$ = 4.0 K in the full-resolution maps}
\end{table}
\newpage

\begin{table}[htb]
\begin{center}
\caption{Parameters of Emission Features}
\begin{tabular}{rrrrrrrr}
\tableline
 Feature & Offset from nucleus & $M_{CO}$\tablenotemark{a} & $\sigma_{1d}\tablenotemark{b}$ & $d_{\alpha}\times d_{\delta}$  &
 $M_{vir}$     & $M_{vir}$ uncertainty\\
     & (arcmin)            & ($10^6 $M$_{\sun}$)            & (km s$^{-1}$)           & (pc) &
 ($10^6$ M$_{\sun}$) & ($10^6$ M$_{\sun}$) \\
\tableline
 1   & (0.65,0.79)       & 2.3       &  6             & 140x150\tablenotemark{c} & $<$2.8 & 1.7 & \\ 
 2   & (0.82,0.76)       & 1.5       &  9             & 90x150\tablenotemark{c} & $<$5.2 & 2.8 & \\ 
 3   & (0.91,0.99)       & 6.6       &  8             & 200x150\tablenotemark{c} & $<$6.1 & 3.4 & \\ 
 4   & (1.03,0.97)       & 5.3       &  9             & 90x150  & 5.2      & 2.8 & \\ 
 5   & (1.08,1.12)       & 2.9       &  6             & 90x110  & 2.0      & 1.2 & \\ 
 6   & (1.54,0.69)       & 16        &  10            & 150x320  & 12       & 6.4 & \\ 
 7   & (1.61,0.42)       & 14        &  9             & 170x380 & 11       & 6.0 & \\ 
 8   & (1.51,0.26)       & 11        &  6             & 150x300  & 4.2      & 2.5 & \\ 
 9   & (1.69,--0.24)       & 7.3       & 6            & 120x270  & 3.6      & 2.2 & \\ 
 10  & (1.48,--0.20)       & 2.0       & 11           & 120x180  & 10       & 5.3 & \\ 
\tableline
\end{tabular}
\end{center}
\tablenotetext{a}{Fractional uncertainty in $M_{CO}$ is 0.8}
\tablenotetext{b}{Typical uncertainty in $\sigma_{1d}$ is 2 km s$^{-1}$}
\tablenotetext{c}{$d_{\delta}$ is an upper limit}
\end{table}

\end{document}